\documentclass[12pt]{article}
\usepackage{graphicx}

\title{Deformed Heisenberg algebra with upper bound of momentum value}
\author{T.V. Fityo\footnote{E-mail: fityo@ktf.franko.lviv.ua}\\
{\small Chair of Theoretical Physics, Ivan Franko National University of Lviv,}\\
{\small 12 Drahomanov St., Lviv, UA-79005, Ukraine}}

\newcommand{\aver}[1]{\left<#1\right>}
\newcommand{\averp}[2]{\left<#2\left|#1\right|#2\right>}

\begin{document}
\maketitle

\abstract{We consider one dimensional deformed Heisenberg algebra
leading to existence of minimal length for coordinate operator and
minimal and maximal uncertainty of momentum operator. For this
algebra an exactly solvable Hamiltonian is constructed.}




\section{Introduction}

Several independent lines for investigation of matter properties
at high energies (string theory \cite{Witten96}, black hole
physics \cite{Maggiore93}, etc., also see \cite{Garay95}) propose
that uncertainty of coordinate $\Delta X$ depends on uncertainty
of momentum $\Delta P$ in such a way
\begin{equation}\label{x1}
\Delta X\ge \frac\hbar2\left(\frac1{\Delta P}+\beta{\Delta
P}\right).
\end{equation}
Minimizing the right part of this expression one obtains that
uncertainty of coordinate is always larger than some threshold
value $\Delta X_{min}=\hbar\sqrt\beta$.

Kempf showed that expression (\ref{x1}) could be derived using
Heisenberg uncertainty inequality from deformed commutation
relation \cite{Kempf94}
\begin{equation}\label{x3}
[X,P]=i\hbar(1+\beta P^2).
\end{equation}
It was shown that there existed states for which $\Delta X<\Delta
X_{min}$, but they are formal states. That is, mean value of
kinetic energy $\aver{P^2/2m}$ does not exist for these states
\cite{Kempf95}. Note, that there exist other algebras
approximately leading to the inequality (\ref{x1})
\cite{Maggiore93b,Quesne06}, but Kempf's algebra is exact, the
simplest and the best studied one.

It is well known that for arbitrary hermitian operators $A$ and
$B$ the Heisenberg inequality holds
\begin{equation}\label{x2}
(\Delta A)^2(\Delta B)^2\ge\frac12\aver{C}^2,
\end{equation}
where $C=[A,B]/i$ is the hermitian operator too. This inequality
holds for all states for which mean values of $A^2$, $B^2$, $AB$
and $BA$ are defined (more rigid conditions of holding of the
inequality (\ref{x2}) may be found in many mathematical papers,
see for instance \cite{Folland97}). Everywhere throughout the
paper $\Delta A$ denotes $\sqrt{\aver{(A-\aver{A})^2}}$.

If someone tries to analyze deformed commutation relation
\begin{equation}\label{x3g}
[X,P]=if(X,P)
\end{equation}
then many interesting properties from corresponding Heisenberg
inequality may arise (e.g., one can discover that  a minimal
length must be present, as for the case of Kempf algebra
(\ref{x3})). In order to investigate such properties one should
consider, for which states the Heisenberg inequality breaks.

We can outline two approaches to such an analysis. The first one
is common and it says that breaking states must be considered
carefully. They may be physically accepted states if mean values
of operators having clear physical sense converge
\cite{Detournay02}. For an example in \cite{Nouicer05} states
breaking corresponding Heisenberg inequality were considered
meaningful since $\aver{P^2}$ converged (it was implicitly assumed
that $\aver{X}$, $\aver{P}$, etc converged too). As it was noted
above, for algebra (\ref{x3}) breaking states are considered
meaningless since mean value of kinetic energy diverges
\cite{Kempf95}.

The second approach says\footnote{This approach was privately
pointed to me by Prof. V.~M.~Tkachuk} that commutation relation
(\ref{x3g}) has physical meaning itself, so Heisenberg inequality
must holds for any physically accepted states. All states breaking
the inequality are considered to be formal and thus having no
physical meaning. In other words, we restrict ourselves to states
satisfying corresponding Heisenberg inequality.

Each algebra (\ref{x3g}) is characterized by several parameters
describing deformation of canonical commutation relation. It is
reasonable to assume that {\em if these parameters tend to zero
then eigenvalues of some system tend to corresponding eigenvalues
of the same system with undeformed commutation relation}. For
algebra (\ref{x3}) all known exactly solvable systems
\cite{Kempf95, Chang02, Quesne05, Fityo06} have this property.
This just formulated correspondence principle can be used for
results verification. We will use it in the third section to
decide which eigenvalues belong to spectrum.

This paper is organized as follows. We describe a deformed
commutation relation leading to upper bound of momentum
uncertainty in the second section. In the third section we write
an exactly solvable Hamiltonian down. And end the paper by several
concluding remarks.

\section{Algebra leading to upper bound of momentum uncertainty}

In order to simplify notation and calculus we use $\hbar=1$,
$m=\frac12$ further in the paper.

In \cite{Quesne06} a large set of different deformed algebras
leading to minimal length was presented. They have the following
structure
\begin{equation}\label{x4}
[f(X),P]=i(f'(X)+\beta P^2).
\end{equation}
Here case of $\beta=0$ corresponds to case of the usual Heisenberg
algebra, where $[X,P]=i$. It was shown that there existed minimal
length for a wide class of such algebras. For details see
\cite{Quesne06}.

Particular form of such a deformed algebra used function
$f(X)=\tanh X$. In this paper we want to present a further
generalization of this algebra presented in \cite{Quesne06}:
\begin{equation}\label{x5}
[\tanh\alpha X,P]=i\left(\frac{\alpha}{\cosh^2\alpha X}+\beta
P^2+\delta\right),\quad \beta>0,\ \delta>0.
\end{equation}
An additional term $\delta$ appears in our generalization. If
someone puts $\beta=0$, $\delta=0$ then an canonical commutation
relation is recovered. For $\alpha X\ll1$ one obtains an algebra
similar to algebra (\ref{x3}).

Let us make two assumptions that operator $\tanh\alpha X$ is a
bounded one and that $\Delta \tanh\alpha X\le 1$ holds for all
states (it would be obvious if $X$ has a complete set of
eigenfunctions). Then applying inequality (\ref{x2}) to
commutation relation (\ref{x5}) one obtains the following chain of
inequalities
\begin{equation}\label{x6}
\Delta P\geq\Delta\tanh\alpha X\Delta P\geq\frac12\left(\Delta
\frac{\alpha}{\cosh^2\alpha X}+ \beta\Delta P^2+\delta\right)
\geq\frac12\left(\beta\Delta P^2+\delta\right).
\end{equation}
From this chain the following constraints on momentum uncertainty
can be derived
\begin{equation}\label{x7}
\frac{\delta}2\leq\Delta P\leq\frac{2}{\beta}.
\end{equation}
In a similar way one can deduce that
\begin{equation}\label{x8}
\frac{\delta^2}4\leq\aver{P^2}\leq\frac 4{\beta^2}.
\end{equation}
Minimizing inequality (\ref{x6}) the following constraints on
coordinate uncertainty appears
\begin{equation}\label{x9}
\sqrt{\beta\delta}\le\Delta\tanh\alpha X\le\alpha\Delta X.
\end{equation}

Existence of lower bounds for uncertainties of momentum and
coordinate operators means that eigenstates of these operators
satisfying inequality (\ref{x6}) do not exist ($\Delta A=0$ for
eigenstates of operator $A$).

The inequality (\ref{x6}) breaks if at least one integral among
$\aver{\tanh^2\alpha X}$, $\aver{P^2}$, $\aver{P\tanh\alpha X}$ or
$\aver{\tanh\alpha XP}$ diverges. Operator $\tanh^2\alpha X$ is a
bounded operator, so the first integral must converge for
normalizable states. Divergence of $\averp{P^2}\psi$ means that
kinetic energy is not well defined in state $\psi$. We could not
analyze the last two integrals separately, but their difference
$\aver{[\tanh\alpha X,P]}$ contains terms proportional to
$\tanh^2\alpha X$, $1$ and $P^2$. The first and the second
operators are bounded ones, the third is kinetic energy operator
again. This fact gives a strong evidence that states breaking
inequality (\ref{x6}) are nonphysical.

Note, that the same situation occurs in the case of deformed
commutation relation (\ref{x3}): mean value of kinetic energy
diverges for states for which inequality (\ref{x1}) breaks
\cite{Kempf95}. Nature of the breaking was established since
explicit representation of $P$ and $X$ operators exists for the
algebra (\ref{x3}). Contrarily, we do not know explicit
representation of $X$ and $P$ operators of the algebra (\ref{x5}).
Only approximate representation can be found in the fashion of
paper \cite{Quesne06} method:
\begin{equation}\label{x10}
X=x, \quad P\approx p+\beta\left(\frac1{6\alpha}\{\cosh^2\alpha
x,4\alpha^2p+p^3\}-p \right) + \frac\delta{2\alpha}\{\cosh^2\alpha
x,p\},
\end{equation}
where small operators $x$ and $p$ satisfy conventional commutation
relation $[x,p]=i$. In representation (\ref{x10}) operators $X$
and $P$ satisfy relation (\ref{x5}) in linear approximation over
parameters $\beta$, $\delta$. Such an approximation is valid far
small values of $x$ and $p$ (when the first term of approximation
(\ref{x10}) is much larger than the second one).


\section{Exactly solvable model}

It is possible to construct exactly solvable Hamiltonian in the
frame of the deformed algebra (\ref{x5}). We use shape-invariance
method to build such a model \cite{Coo95}. Let us introduce
annihilation-creation operators
\begin{equation}\label{x20}
A_n=i\xi_n P+\eta_n\tanh X,
\end{equation}
\begin{equation}\label{x21}
A^+_n=-i\xi_n P+\eta_n\tanh X.
\end{equation}
Note that we also fix $\alpha=1$ in algebra (\ref{x5}).

On the basis of these operators we build partner Hamiltonians
\begin{equation}\label{x22}
H^-_n=A^+_nA_n=(\xi^2_n-\xi_n\eta_n\beta)P^2-\frac{\eta^2_n+\xi_n\eta_n}{\cosh^2X}
+ \eta_n^2-\xi_n\eta_n\delta,
\end{equation}
\begin{equation}\label{x23}
H^+_n=A_nA^+_n=(\xi_n^2+\xi_n\eta_n\beta)P^2-\frac{\eta^2_n-\xi_n\eta_n}{\cosh^2
X} + \eta_n^2+\xi_n\eta_n\delta.
\end{equation}
These Hamiltonians have similar form to the so-called
P\"oschl-Teller Hamiltonian $H=p^2-1/\cosh^2x$ which is an exactly
solvable one \cite{Coo95} in undeformed case.

To find spectrum of Hamiltonian $H_0^-$ we construct a chain of
such partners:
\begin{equation}\label{x24}
H^+_{n-1}=H^-_n+\epsilon_n,
\end{equation}
where $\epsilon_n$ is a constant. The following connections
between parameters $\xi$, $\eta$ and $\epsilon$ can be deduced
from equation (\ref{x24})
\begin{equation}\label{x25}
\xi_n=\frac{\xi_{n-1}+\beta\eta_{n-1}}{\sqrt{1+\beta}}, \
\eta_n=\frac{\eta_{n-1}-\xi_{n-1}}{\sqrt{1+\beta}}, \
\epsilon_n=(\eta_{n-1}^2-\eta_n^2)(1+\delta).
\end{equation}
Eigenvalues of Hamiltonian $H_0^-$ read
\begin{equation}\label{x26}
E_n=\sum_{i=1}^n\epsilon_i=(\eta^2_0-\eta^2_n)(1+\delta).
\end{equation}
$\eta_n$ can be expressed with the help of parameters $\eta_0$,
$\xi_0$ of initial Hamiltonian $H^-_0$
\begin{equation}\label{x27}
\eta_n=\eta_0\cos n\theta-\xi_0\frac{1}{\sqrt{\beta}}\sin n\theta,
\end{equation}
where $\cos\theta=\frac{1}{\sqrt{1+\beta}}$ and
$\sin\theta=\sqrt{\frac{\beta}{1+\beta}}$.

Integer quantum number $n$ varies in the range from $0$ to
$n_{max}$, where $n_{max}$ is the greatest integer less than
$\frac1\theta \arctan\frac{\eta_0\sqrt\beta} {\xi_0}$. This
condition was derived from the requirement that
\begin{equation}\label{x28}
\eta_n>0.
\end{equation}
If this requirement breaks then the signs in the expression for
annihilation operator (\ref{x20}) changes. Here we appeal to the
correspondence principle formulated in the Introduction: if
$\eta_n$ changes sign then $n^{\rm th}$ eigenfunction becomes
unnormalizable in the undeformed space. In deformed case for
$n>n_{max}$ eigenvalues decreases if $n$ increases: it is unusual
feature of quantum systems.



\begin{figure}[htb!]
\centerline{\includegraphics[width=0.7\textwidth,clip]{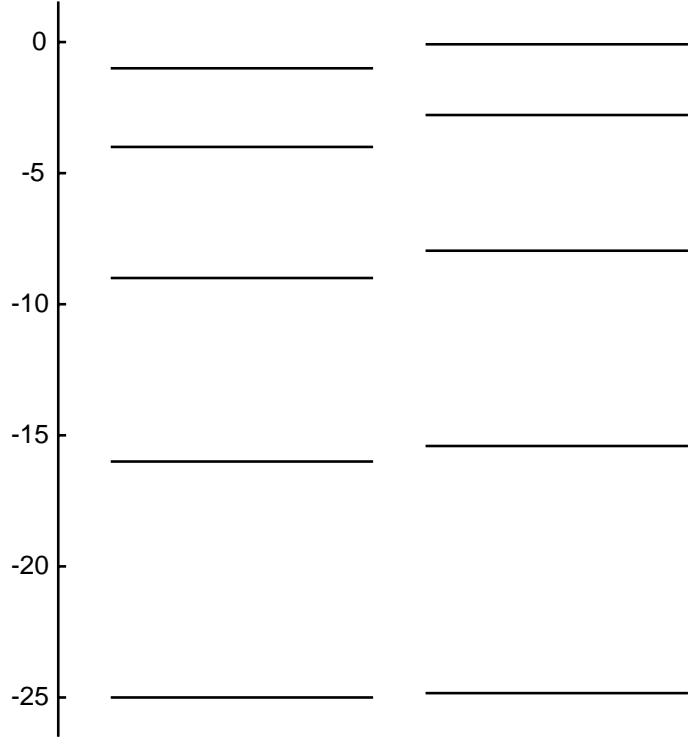}}
\caption{Eigenvalues of Hamiltonian (\ref{x29}) in undeformed case
\cite{Coo95} (left column) and in deformed case: equation
(\ref{x26}); $\beta=0.01$, $\delta=0.01$ (right
column).}\label{SpectraTh2.eps}
\end{figure}

The spectrum of Hamiltonian
\begin{equation}\label{x29}
H=P^2-\frac{30}{\cosh^2X}
\end{equation}
is showed on Fig. \ref{SpectraTh2.eps}. All levels of deformed
problem are above corresponding levels of undeformed one.

In linear approximation over $\beta$, $\delta$ expression
(\ref{x26}) can be simplified to
\begin{equation}\label{x30}
E_n\approx (1+\delta)(\eta_0^2-(\eta_0-\xi_0 n)^2)- \frac\beta3
(\eta_0-\xi_0 n) \left( \xi_0n^3-3\eta_0n^2+2\xi_0n\right).
\end{equation}

Hamiltonian in linear approximation is $$
H_0^-\approx\xi^2_0p^2-\frac{\eta^2_0+\xi_0\eta_0}{\cosh^2x}+\eta_0^2+
\delta\left(\xi_0^2\left\{p,\frac12\{p,\cosh^2x\}\right\}-\xi_0\eta_0\right)+
$$
\begin{equation}\label{x31}
+\beta\left( \xi_0^2 \left\{p,\frac16\{\cosh^2x,4p+p^3\}\right\}
-(2\xi_0^2+\xi_0\eta_0)p^2 \right)=H^{lin}.
\end{equation}
As one should expect $\averp{H^{lin}}{\psi_n}$ exactly coincides
with expressions (\ref{x30}) for all $n$, except for $n=n_{max}$.
Here $\psi_n$ denotes $n^{\rm th}$ eigenfunction of undeformed
P\"oschl-Teller system. This exception arises since corresponding
integral diverges for large $x$, where approximation (\ref{x10})
becomes invalid. But the question on the range of $n$ variation
needs more attention and it seems it can be resolved finally only
if exact representation of algebra (\ref{x5}) is found.

\section{Concluding remarks}

Algebra (\ref{x5}) is not a unique algebra characterizing by upper
bound for mean values of $P^2$ operator. There are many other such
algebras. One of them reads
\begin{equation}\label{x40}
[X,P]=i(1+\alpha X^4+\beta P^2).
\end{equation}
Using Heisenberg inequality (\ref{x2}) it is easy to show that
\begin{equation}\label{x41}
\frac{16}{\alpha\beta^3}\ge\aver{P^2}\ge(\Delta P)^2
\ge\frac{4}{9}(3\alpha)^{\frac{1}{2}},
\end{equation}
\begin{equation}\label{x42}
\frac{4}{\alpha\beta}\ge\aver{X^2}\ge (\Delta X)^2\ge\beta.
\end{equation}
We have not found explicit representation of $X$ and $P$ operators
for it as for algebra (\ref{x5}). We choose algebra (\ref{x5}) for
our analysis since exactly solvable model exists for it and right
side of (\ref{x5}) contains bounded operators and $P^2$ operator
having a clear physical sense of kinetic energy.

Existence of $X$ and $P$ representation is still an open question.
On one hand, in classical mechanics we always can change variables
$(X,P)$ into $(x,p)$ that Poisson bracket $\{x,p\}=1$ (Darboux
theorem, \cite{Arnold89}), for which obvious representation
exists. On the other hand, if $\beta\delta>4$ then algebra
(\ref{x5}) is self-contradictory (as it follows from constraints
(\ref{x7})). It follows from (\ref{x42}) that if $\alpha\beta^2>4$
then algebra (\ref{x40}) is self-contradictory too. Saying that
the algebra is self-contradictory we imply that for such values of
deformation parameters  there are no states satisfying
corresponding Heisenberg inequality.

There are some other open questions which are needed to be
answered. They are: generalization of such complicated algebras as
(\ref{x5}) and (\ref{x40}) to multidimensional case; construction
of analogue of Galilei transformation. It is the third open
question. Obviously one can answer them (partly or completely) if
one finds an explicit representation.


\end{document}